\documentclass[iop]{emulateapj}
\usepackage{comment}
\usepackage{color}
\usepackage{xspace}

\newcommand\um      {\ifmmode\mu{\rm m}\else$\mu${\rm m}\fi\xspace}

\newcommand{\fuv}{{\sl FUV }}
\newcommand{\nuv}{{\sl NUV }}

\slugcomment{Submitted to ApJ \today}
\shortauthors{}

\begin{document}

\title{The Post-Starburst Evolution of Tidal Disruption Event Host Galaxies}

\author{
K. Decker French     \altaffilmark{1},
Iair Arcavi \altaffilmark{2} \altaffilmark{$\dagger$},
Ann Zabludoff      \altaffilmark{1}}

\altaffiltext{1}{Steward Observatory, University of Arizona, 933 North Cherry Avenue, Tucson AZ 85721}
\altaffiltext{2}{Department of Physics, University of California, Santa Barbara, CA 93106-9530, USA}
\altaffiltext{$\dagger$}{Einstein Fellow}

\begin{abstract}
We constrain the recent star formation histories of the host galaxies of eight optical/UV-detected tidal disruption events (TDEs). Six hosts had quick starbursts of $<$ 200 Myr duration that ended 10 to 1000 Myr ago, indicating that TDEs arise at different times in their hostÕs post-starburst evolution.  If the disrupted star formed in the burst or before, the post-burst age constrains its mass, generally excluding O, most B, and highly massive A stars. If the starburst arose from a galaxy merger, the time since the starburst began limits the coalescence timescale and thus the merger mass ratio to more equal than 12:1 in most hosts. This uncommon ratio, if also that of the central supermassive black hole (SMBH) binary, disfavors the scenario in which the TDE rate is boosted by the binary but is insensitive to its mass ratio. The stellar mass fraction created in the burst is $0.5 - 10$\% for most hosts, not enough to explain the observed $30-200\times$ boost in TDE rates, suggesting that the host's core stellar concentration is more important. TDE hosts have stellar masses $10^{9.4}-10^{10.3}$ M$_\sun$, consistent with the SDSS volume-corrected, quiescent Balmer-strong comparison sample and implying SMBH masses of $10^{5.5} - 10^{7.5}$ M$_\sun$. Subtracting the host absorption line spectrum, we uncover emission lines; at least five hosts have ionization sources inconsistent with star formation that instead may be related to circumnuclear gas, merger shocks, or post-AGB stars.

\end{abstract}


%----------------------------------------------------------------------
\section{Introduction}

Tidal Disruption Events (TDEs) are observed when a star passes close enough to a black hole that the tidal forces exceed the self-gravity of the star. Disruptions occurring outside the event horizon (typically for black hole masses $\lesssim10^8M_{\odot}$) are expected to be accompanied by an observable flare \citep{Hills1975,Rees1988, Evans1989, Phinney1989}. Rapid identification and followup enabled by transient surveys have produced a steadily increasing list of TDE candidate events, some detected primarily in X-ray/$\gamma$-ray emission \citep{Bloom2011, Burrows2011, Levan2011, Zauderer2011, BradleyCenko2012, Brown2015} and others in the optical/UV \citep[Blagorodnova et al., in prep]{Gezari2012, Arcavi2014, Holoien2014, Holoien2015, Prentice2015}. 

Curiously, the optical/UV class of TDEs preferentially occur in quiescent galaxies with strong Balmer absorption lines, indicating an intense period of star formation that has recently ended, which was likely a starburst \citep{Arcavi2014, French2016}. Several possibilities have been proposed to explain how the global star-formation history could be connected to the TDE rate. Many post-starburst galaxies show signs of a recent galaxy-galaxy mergers and have centrally concentrated young stellar populations \citep{Zabludoff1996, Yang2004, Yang2008, Swinbank2012}. If TDE host galaxies are post-starburst and post-merger, the TDE rate could be boosted by 1) 3-body interactions with the central supermassive black hole binary, 2) a high concentration of young stars in center-crossing orbits, or 3) residual circumnuclear gas. 

In the sample of eight TDE hosts analyzed by \citet{French2016}, we identified three that were quiescent and had Balmer absorption consistent with post-starburst galaxies (H$\delta_{\rm A}$ $-$ $\sigma$(H$\delta_{\rm A}$) $>$ 4\,\AA, where $\sigma$(H$\delta_{\rm A}$) is the error in the Lick H$\delta_{\rm A}$ index, and H$\alpha$ emission EW $<$ $3$ \AA). Another three were identified as being just ``quiescent Balmer-strong," with H$\delta_{\rm A} >$ 1.31\,\AA\ and H$\alpha$ EW $<$ $3$\, \AA. While these features imply an unusual recent star formation history, we could not differentiate between a recent short period of star formation (a true burst) and longer declines in the star formation rate, nor between older, stronger bursts and younger, weaker bursts.

In this paper, we use stellar population fitting to characterize the recent star-formation histories (SFHs) of TDE host galaxies. We consider the same sample of optical/UV - detected TDEs as in \citet{French2016}. Using a new stellar population fitting method (French et al. 2016, submitted), we break the degeneracy between the duration of recent star formation, the time since its end, and the mass of stars it produced, to determine whether these galaxies have experienced a recent starburst and to use the burst characteristics to constrain possible mechanisms for enhancing the TDE rate. We also determine the galaxy's stellar mass. Our method is more accurate than those assuming a single stellar population, as a recent starburst will impact the mass-to-light ratio. These stellar masses are used to infer black hole masses, using measurements of the bulge mass and the black hole -- bulge mass relation. In addition, we use the stellar population fitting results to uncover hidden emission lines, discriminating between ionization from star formation and from other sources. We compare these TDE host properties to the general quiescent Balmer-strong population (13749 galaxies) from the Sloan Digital Sky Survey (SDSS).

\section{Data}

We use optical photometry from the SDSS, UV photometry from {\it GALEX} and {\it Swift} UVOT, and optical spectroscopy to fit to stellar population models. For the SDSS photometry, we adopt the {\tt modelmag} magnitudes, which provide stable colors while containing most of the galaxy light \citep{Al.2004}. We make small corrections to the $u$ and $z$ bands ($-$0.04 and 0.02 mag) to put them on the correct AB magnitude system. In addition to photometry errors given in the SDSS catalog, we add the magnitude zero-point errors (5\%, 2\%, 2\%, 2\%, and 3\% in $ugriz$) in quadrature.

For each of the TDE host galaxies, we search for matching {\it GALEX} \nuv and \fuv detections using the {\it GALEX} GCAT catalog. We search for galaxies within 4\arcsec\ of the SDSS positions. This radius is similar to the FWHM of the \nuv PSFs and much larger than the {\it GALEX} astrometric uncertainties (0.59\arcsec\ in {\sl FUV}). We have detections or upper limits in the \fuv for the hosts of SDSS J0748, ASASSN-14li, and PTF09ge, and \nuv detections or upper limits for all but the PTF09axc host. We obtain additional archival {\it Swift} UVOT data for the hosts of ASASSN-14ae, PTF09axc, PTF09djl, and PTF09ge. 

We use optical spectra for the hosts of ASASSN-14ae, ASASSN-14li, and PTF15af from the SDSS \citep{Aihara2011}, for SDSS J0748 from \citet{Yang2013}, and the rest from \citet{Arcavi2014}. 

In \citet{French2016}, we also consider the high energy TDE Swift J1644. We do not present results for its host galaxy in this work, due to the absence of rest-frame $FUV$ or $NUV$ photometry, as well as the absence of a medium-resolution spectrum (R$\sim$1000), required for our age-dating method.

We compare the TDE host galaxy properties with galaxies from the SDSS main galaxy spectroscopic sample \citep{Strauss2002} DR10 \citep{Aihara2011}. As in \citet{French2016}, we select quiescent, Balmer-strong galaxies to have H$\delta_{\rm A} >$ 1.31\,\AA\ and H$\alpha$ EW $<$ $3$\, \AA.

\section{Methods: Age-Dating Host Galaxies}
To constrain the detailed star formation histories of the TDE host galaxies, we fit the spectroscopic and photometric properties of these galaxies using stellar population models. The method is described and tested fully in (French et al. 2016, submitted), We describe it briefly here. 

We use the flexible stellar population synthesis (FSPS) models of \citet{Conroy2009} and \citet{Conroy2010} to construct model template spectra. We assume a Chabrier IMF \citep{Chabrier2003} and a Calzetti reddenning curve \citep{Calzetti2000}. We model the spectra of quiescent Balmer-strong galaxies $f_{model}$ as a combination of old and young stellar populations. The old stellar population's star formation history is modeled as a linear-exponential star formation rate over time $t_{old}$:
\begin{equation}
\Psi \propto t_{old} e^{-t_{old}/\tau_{old}}% ; \ \tau_{old} = 1 \ \rm{Gyr}
\label{eqn:old}
\end{equation}
beginning 10 Gyr ago and characterized by the timescale $\tau_{old} = 1$ Gyr. 
The young stellar population's star formation history is modeled as an exponential decline in the star formation rate over time $t_{young}$: 
\begin{equation}
\Psi \propto e^{-t_{young}/\tau}.% ; \ \tau = 0.025-3 \ \rm{Gyr},
\label{eqn:young}
\end{equation}
We vary the time $t_{SB}$ since this recent period of star formation began\footnote{Equations 1 and 2 are related by $t_{old}-t_{young} = 10$ Gyr $- t_{SB}$.} as well as the characteristic timescale $\tau$. The observed spectrum is modeled as a linear combination of the young and old stellar templates:
\begin{equation}
f_{model}  = [y f_{young} + (1-y) f_{old}] \times 10^{-0.4 k(\lambda) A_V},
\end{equation}
where $k(\lambda)$ is the reddening curve as a function of the wavelength $\lambda$, $A_V$ is the amount of
internal extinction expressed in magnitudes of $V$-band absorption, $f_{young} (\lambda; t_{SB}, \tau, Z)$ is the young stellar population spectrum (arising from Eq. \ref{eqn:young}), and $f_{old}(\lambda; Z)$ is the old stellar population spectrum (arising from Eq. \ref{eqn:old}). $Z$ is the stellar metallicity assumed, using the priors described below. Each spectrum is normalized within the 5200--5800 \AA\ wavelength window, and $y$ represents the fraction of the total galaxy light in the young stellar template. The mass fraction of new stars $m_f$ is derived from $y$ and $t_{SB}$ after the fitting is complete. Thus, we parameterize the spectrum using four free parameters, $t_{SB}$, $A_V$, $y$, and $\tau$. The priors on these four parameters are: $A_V$, [0, 5] magnitudes, spaced linearly; $t_{SB}$, [30, 2000] Myr, spaced logarithmically; $y$, [0.01, 1], spaced logarithmically; $\tau$, [25, 50, 100, 150, 200, 1000, 3000] Myr.

We compare the SDSS $ugriz$ and {\it GALEX} $FUV, NUV$ photometry, as well as 23 Lick indices \citep{Worthey1994, Worthey1997}, to synthetic photometry and Lick indices calculated from the model spectra. The UV photometry is especially sensitive to the age and duration of the young stellar population, and with the Lick indices, which carry information about the young and old stellar populations, allows us to break the degeneracies between the age, duration, and mass fraction of the young stellar population. We determine the best fit model using $\chi^2$ minimization, and marginalize over all other parameters to determine the 68\% likelihoods for each parameter. Errors on the new stellar mass fraction are determined using a Monte Carlo method to propagate the errors on the age and new stellar light fraction. 

We perform this fitting procedure twice. The first iteration, we assume solar metallicity. Using the stellar mass fit during the first iteration, we use the predicted metallicity from the mass-metallicity relation from \citet{Gallazzi2005} in the second iteration. 

One consideration is the possibility of contamination from the current TDE or past TDEs. The {\it GALEX} photometry for each galaxy used here was taken prior to the current TDE, while the {\it Swift} photometry is post-TDE and thus may be contaminated.  However, most of the {\it Swift} data are upper-limits, and their removal does not change our derived parameters more than the quoted errors.

If there were a recent TDE prior to the current one, the {\it GALEX} data could be contaminated. For the six quiescent Balmer-strong TDE host galaxies, the expected TDE rate is 2-4$\times10^{-4}$ year$^{-1}$ per galaxy \citep{French2016}. The enhanced UV emission thus would need to persist for $>$400 years to affect our results:  in that case, the calculated post-burst age of at least one of the hosts would be underestimated beyond the quoted errors. Even for the most TDE-active galaxies, with a TDE rate of $3\times10^{-3}$ year$^{-1}$ per galaxy, the enhanced UV emission must persist for $>$125 years to significantly alter at least one host galaxy's UV data. 

However, the long-term UV-optical evolution of TDEs ASASSN-14ae and ASASSN-14li \citep{Brown2016,Brown2016b} shows the emission declining rapidly over the first 100 {\it days} after the event, leveling off after 300 days. The optical emission, including the U-band flux, returns to that expected for the host. The UV light-curve declines over the same timescale, but levels off at brighter magnitudes than expected.  We note that the UV spectra of post-starburst galaxies are not well-fit by single stellar populations, which may explain the discrepancies between the predicted final UV and optical fluxes for the host.  Were the final host UV excess to be real, and to persist at 0.5-1 mag over the next 400 years, we would underestimate the post-burst age for at least one galaxy in our sample. That is unlikely; assuming a decline of $t^{-5/3}$ and the characteristic timescale of two months from \citet{Holoien2015}, the TDE flux would drop by a factor of 60,000 in 125 years. Even for a shallower power law decline of $t^-{5/12}$ (as expected from disk emission at late times; \citealt{Lodato2011}), the TDE flux would drop by $>15\times$ in 125 years, more than enough to be consistent with the host galaxy.

\begin{table*}[!t]
\centering
\caption{TDE Host Galaxy Star Formation Histories}
\label{table:sfh}
\begin{tabular}{l r r r r}
\hline
\hline
TDE &  Starburst Age$^a$ & Post-Starburst Age$^b$ & Burst Mass Fraction $m_f^c$ & Burst Duration $\tau^d$ \\
 & (Myr) & (Myr) & & (Myr)  \\
\hline
SDSS J0748 &  354$^{+ 177} _{- 176}$  & -1150$^{+ 177} _{-  176}$  & 0.005 $^{+  0.001} _{- 0.001 }$ &1000 (50-1000)  \\
ASASSN14ae$^\dagger$ &  501$^{+  37} _{-  82}$  &   155$^{+  37} _{-   82}$  & 0.019 $^{+  0.004} _{- 0.004 }$ & 150 (100-150)  \\
ASASSN14li$^\dagger$ &  473$^{+ 373} _{-  67}$  &   415$^{+ 373} _{-   67}$  & 0.055 $^{+  0.016} _{- 0.016 }$ &  25 (25-200)  \\
PTF09axc & 1496$^{+ 155} _{- 155}$  &  1036$^{+ 155} _{-  155}$  & 0.890 $^{+0.110} _{- 0.211 }$ & 200  (25-200)  \\
PTF09djl &  211$^{+  59} _{-  23}$  &   153$^{+  59} _{-   23}$  & 0.023 $^{+0.005} _{- 0.005 }$ &  25 (25-200)  \\
PTF09ge &  530$^{+ 730} _{-  69}$  &   415$^{+ 730} _{-   69}$  & 0.009 $^{+0.012} _{- 0.012 }$ &  25 (25-200)  \\
PS1-10jh &  118$^{+ 722} _{-  15}$  &    61$^{+ 722} _{-   15}$  & 0.006 $^{+0.002} _{- 0.002 }$ &1000 (25-1000)  \\
PTF15af$^\dagger$ &  595$^{+ 108} _{- 191}$  &   538$^{+ 108} _{-  191}$  & 0.005 $^{+0.002} _{- 0.002 }$ &  25 (25-100)  \\
\hline

\end{tabular}
\begin{flushleft}
Notes: (a) Age since starburst began ($t_{SB}$), with 1$\sigma$ errors. (b) Age since 90\% of the young stars formed, or post-burst age, with 1$\sigma$ errors. (c) Fraction of current stellar mass produced in the recent period of star formation, with 1$\sigma$ errors. (d) Characteristic exponential timescale of the recent period of star formation, shown with 90\% error bounds due to the coarseness of the fitting grid used and large uncertainties due to the small burst mass fractions. $^\dagger$ TDE hosts for which the host spectra were obtained before the TDE.
\end{flushleft}
\end{table*}

\begin{table*}[!t]
\centering
\caption{TDE Host Galaxy Properties}
\label{table:sfh2}
\begin{tabular}{l r r r r r r r r }
\hline
\hline
\small
TDE &  $A_V$ & M$_\star^a$ & M$_{\rm bulge}^b$ & M$_{BH}^c$ &[NII]/H$\alpha$ & [OIII]/H$\beta$ & H$\alpha$ EW$^d$   \\
 & (mag)  & ----- & $(log$ M$_\sun)$ & -----  & & & (\AA)  \\
\hline
SDSS J0748 &  $<0.1$ & 10.2  & 8.1 & 5.5  & 0.31$\pm$0.03 & 1.07$\pm$0.10 & -11.36$\pm$1.00   \\
ASASSN14ae$^\dagger$ &   $<0.1$ & 9.9  & 9.6 & 6.9 & 0.55$\pm$0.17 & 6.28$\pm$2.16 & -0.68$\pm$0.40   \\
ASASSN14li$^\dagger$ &    0.3 & 9.7 &  9.6 & 6.9  & 0.89$\pm$0.14& $<12.5$ & -0.59$\pm$0.53   \\
PTF09axc &  $<0.1$ & 9.7 &  & & 0.90$\pm$0.14 & & -1.07$\pm$0.67    \\
PTF09djl  & 0.4 & 10.2 &  & & $<3.96$ & $<3.4$ & -0.26$\pm$0.66   \\
PTF09ge  &  $<0.1$ & 10.2 &  10.0  & 7.5 & 0.80$\pm$0.04 & 3.41$\pm$0.93 & -1.70$\pm$0.75   \\
PS1-10jh  &  $<0.1$ & 9.4 & &  & 0.58$\pm$0.21 & & -0.54$\pm$0.65   \\
PTF15af$^\dagger$  & 0.3 & 10.3 & &  & $<0.62$ & & -1.65$\pm$0.30   \\
\hline

\end{tabular}
\begin{flushleft}
Notes: (a) Log Stellar Mass, typical error 0.1 dex. (b) Log Bulge Mass, scaled from stellar mass, and bulge mass fraction from \citet{Mendel2013}, typical error 0.25 dex. (c) Scaled from bulge mass and \citet{McConnell2012} relation. The intrinsic scatter ($0.3-0.4$ dex) in this relation will dominate the errors. (d) Same as in \citet{French2016}. $^\dagger$ TDE hosts for which the host spectra were obtained before the TDE.
\end{flushleft}
\end{table*}

\section{Results and Discussion}

\subsection{Star Formation Histories of TDE Host Galaxies}

\subsubsection{Evidence for Starbursts}

With the fitted SFH information, we can distinguish whether the TDE hosts have experienced a true starburst ($\lesssim$200 Myr) or just a recent decline in a longer period of star formation. We allow for a range of possible durations for the recent star formation episode, from $\tau=$ 25 Myr to 3 Gyr. Best-fit durations are shown for each galaxy in Table \ref{table:sfh}. Only two TDE hosts prefer a burst duration of 1 Gyr, longer than expected from a true starburst: SDSS J0748, which is still forming stars (so $\tau$ is less meaningful) and PS1-10jh. The other six TDE hosts are truly post-starburst, including PTF09ge, despite its weaker Balmer absorption \citep{French2016}. The fraction of TDE hosts at each burst duration is consistent with the number expected from the comparison sample of quiescent Balmer-strong SDSS galaxies.

\subsubsection{Post-Starburst Ages}

We define two different ages for these galaxies: the age since the starburst began ($t_{SB}$) and the post-burst age (the age since the starburst ``ended," when 90\% of the stars in the recent burst had formed). Both sets of derived ages are shown in Table \ref{table:sfh}. These two age definitions differ depending on the burst duration $\tau$. The youngest age is for the host of SDSS J0748, which is still star-forming and thus has a negative ``post-burst age." The oldest is the host of PTF09axc, with a post-burst age of 1 Gyr. The range in ages is physical, and implies that TDEs are not limited to a specific point in their hostsÕ post-starburst evolution.

We compare the derived properties of the TDE hosts with the SDSS quiescent Balmer-strong galaxies (Figure \ref{fig:age}a). The shape of the SDSS sample contours is set by when H$\delta$ absorption decreases (at later ages for stronger bursts) on the right hand side, and by a combination of effects on the left hand side. Because galaxies with SFR $\gtrsim 300$ M$_\sun$ yr$^{-1}$ rarely exist in the local universe, starbursts with high burst mass fractions must form stars over a longer duration, delaying their entry into our selection criteria. The TDE hosts, however, are not subject to these selection effects, so the absence of TDE hosts at long post-burst ages is physical.

When comparing the six quiescent Balmer-strong TDE hosts to the SDSS sample, there is a slight dearth of TDE hosts at post burst ages $>0.6$ Gyr (five have post-burst ages $<0.6$ Gyr and only one has a greater post-burst age). In contrast, 53\% of the quiescent Balmer-strong SDSS galaxies have post-burst ages $<0.6$ Gyr. Given the small numbers, the binomial confidence interval for the TDE host galaxies is 0.028 - 0.45, making the significance of this difference only 1.3$\sigma$. This difference persists if we compare the ages since the starburst began, instead of the ages since the starburst ended (which will differ depending on the burst duration). More TDEs are needed to establish whether they are preferentially found in younger post-burst galaxies.

\subsubsection{Starburst Mass Fractions}

For 7/8 of the TDE hosts, the stellar mass of the young stellar population is $0.5-10$\% of the total stellar mass. These burst mass fractions are consistent with those for the SDSS quiescent Balmer-strong galaxies; the distribution about 10\% burst mass fraction is not significantly different between the two samples. These burst mass fractions are typically lower than those found in galaxies with stronger Balmer absorption (H$\delta_{\rm A}$ $-$ $\sigma$(H$\delta_{\rm A}$) $>$ 4\,\AA), which have burst mass fractions of $3-50$\%. Thus, the lower burst mass fractions (not longer duration periods of star formation nor more time elapsed since the starburst ended) are responsible for the weaker Balmer absorption seen in some of the TDE hosts. 

As a comparison, normal star-forming galaxies with similar stellar masses as the TDE hosts have SFRs $\sim$1 M$_\sun$yr$^{-1}$ \citep{Brinchmann2004}, and so would form $0.25-2$\% of their current stellar mass over the same 25-200 Myr period. While the TDE host burst fractions are typically higher, they do not produce enough new stars to account for the  $30-200\times$ boost \citep{French2016} in their TDE rates, relative to normal star-forming galaxies. If the number of newly formed stars is not the driver, the concentration of the stars in the core must be more important. More work is needed to determine if the stellar concentration near the nucleus is sufficient to explain the boosted TDE rates in these galaxies (e.g., \citealt{Stone2016}).

\subsubsection{Dust Extinction}

In Table \ref{table:sfh2}, we show the dust extinctions best fit to the spectra. The extinctions are low, $A_V < 0.4$ mag, consistent with those best fit in the SDSS quiescent Balmer-strong sample. In comparison, the extinctions fit by the MPA JHU group \citep{Brinchmann2004} for early type galaxies (selected using a cut on H$\alpha$ EW $<$ $3$\, \AA) with similar stellar masses ($10^{9.5}-10^{10.5}$) are similar, with a median of  $A_V = 0.28$ mag. Therefore, it is unlikely that the TDE preference for quiescent Balmer-strong hosts arises from easier detection, as the extinction tends to be comparable in early type galaxies where fewer TDEs are found.

\subsubsection{Constraints on the TDE Progenitor Star}
\label{sec:startiming}

Using the post-burst ages, we can constrain the likely mass of the star that was tidally disrupted, assuming it formed in the recent burst or before. This constraint is independent of previous ones, which modeled the mass accreted during the TDE itself \citep[e.g.,][]{Guillochon2014a}. We plot the post-burst ages vs. burst mass fractions in Figure \ref{fig:age}a.  On the top axis, we compare the post-burst ages to the stellar lifetimes of stars created near the end of the starburst. Stars more massive than these will have evolved off the main sequence by the time the TDE occurred. The range in post-burst ages is much larger than the expected post-main sequence evolution timescales, excluding a specific phase of post-main sequence evolution as the cause of the enhanced TDE rate after the starburst. Additionally, it is unlikely that the disrupted stars were giants evolving off the main sequence, because such stars are expected to be disrupted in multiple epochs, resulting in lower luminosity, longer duration TDEs than what was observed for these events \citep{Macleod2012, Macleod2013}. Therefore, we assume that the progenitor was a main sequence star and examine the implications.

We cannot place constraints on the star disrupted in SDSS J0748, as the host galaxy is still actively forming stars. For the seven quiescent hosts, the upper limits on the mass of the likely disrupted star range widely from $\sim$2.5, $\sim$4, $\sim$6, to $\sim$9 M$_\sun$. As a consequence, O stars, as well as most B and highly massive A stars, are excluded as TDE progenitors. Observations of TDEs in older post-starburst galaxies would place stronger constraints on the likely progenitor mass for the disrupted star in those hosts.

\subsubsection{Constraints on the Black Hole Binary Mass Ratio}
\label{sec:mergertiming}
\begin{figure*}
\includegraphics[width=0.9\textwidth]{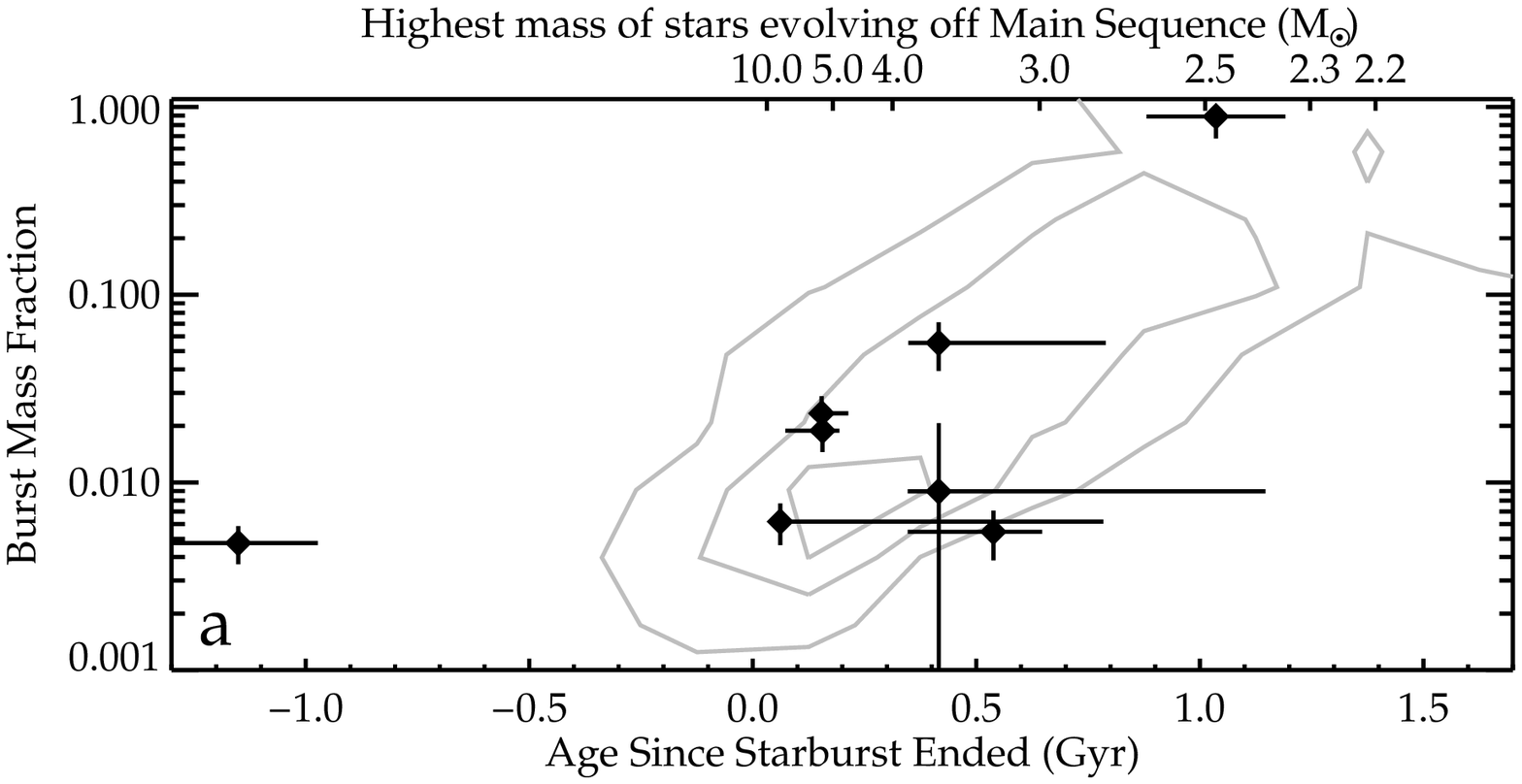}
\includegraphics[width=0.9\textwidth]{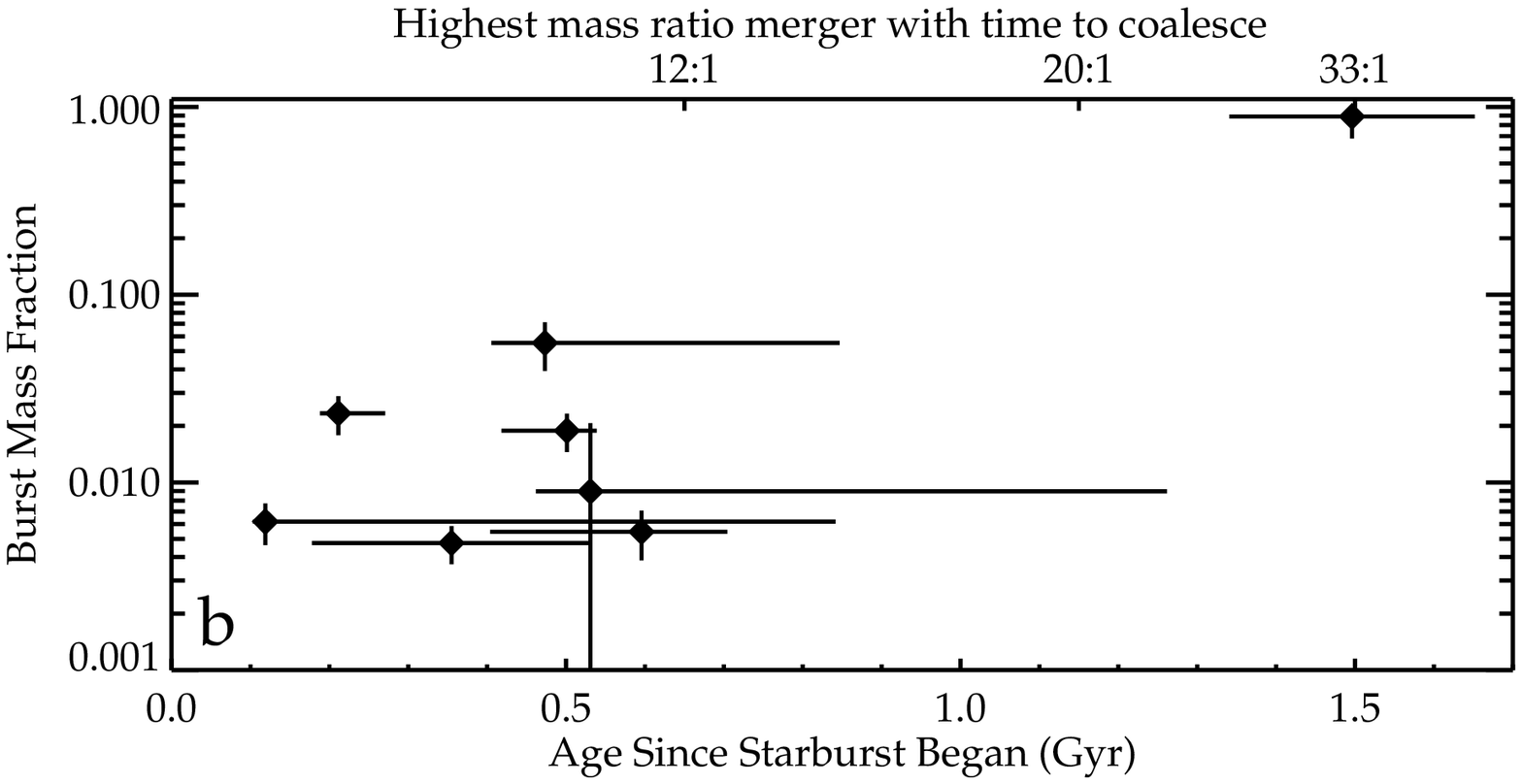}
\caption{{\bf a.} Post-burst age vs. burst mass fraction for TDE hosts and SDSS quiescent Balmer-strong comparison sample.  There is a physical spread in the post-burst ages of the TDE hosts. The youngest galaxy (host of SDSS J0748) is still star-forming, and so has a ``negative" post-burst age. The oldest is the host of PTF09axc, with a post-burst age of 1 Gyr.  25\%, 68\% and 95\% contours are shown for the quiescent Balmer-strong (H$\delta_{\rm A} >$ 1.31\,\AA) sample. The TDE hosts have ages and burst mass fractions consistent with this sample. The shape of the SDSS sample contours is set by when H$\delta$ absorption decreases (at later ages for stronger bursts) on the right hand side, and by a combination of effects on the left hand side. Because galaxies with SFR$\gtrsim 300$ M$_\sun$ yr$^{-1}$ rarely exist in the local universe, starbursts with high burst mass fractions must form stars over a longer duration, delaying their entry into our selection criteria. The TDE hosts, however, are not subject to these selection effects, and the absence of TDE hosts at long post-burst ages is physical.
The top x-axis shows main sequence lifetimes corresponding to the highest mass stars that have not yet evolved off the main sequence for each post-burst age. The star disrupted in PTF09axc likely had a mass of M $\lesssim 2.5$M$_\sun$. We cannot place constraints on the star disrupted in SDSS J0748, as the host galaxy is still actively forming stars. For the other host galaxies considered here, the constraints on the mass of the disrupted star range from M $\lesssim 3-10$M$_\sun$, ruling out O, B, and the most massive A stars as likely disrupted stars.
{\bf b.} Age since the starburst began vs. burst mass fraction for TDE hosts, with the most unequal mass galaxy merger that could have coalesced via dynamical friction. All of the hosts but that of PTF09axc are consistent with a merger mass ratio more equal than $12:1$.  If supermassive black hole binaries were driving the TDE enhancement, and if the TDE rate enhancement were insensitive to the SMBH binary mass ratio \citep{Chen2011}, we would expect more unequal mass ratio mergers, since these are more common.}
\label{fig:age}
\end{figure*}

\begin{figure}
\includegraphics[width=0.49\textwidth]{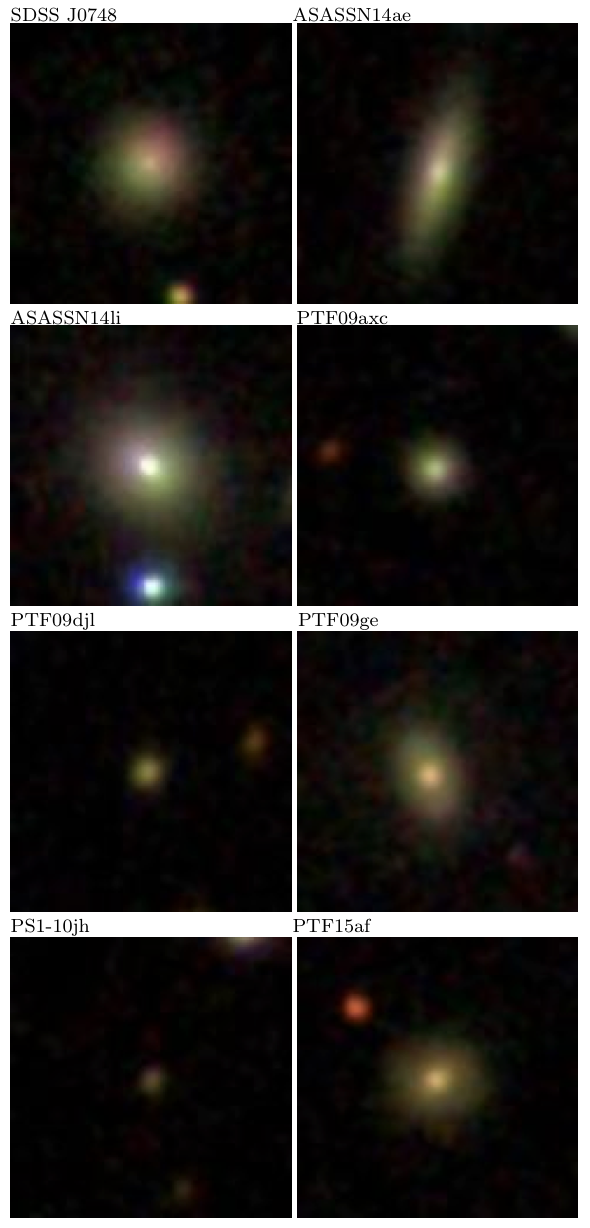}
\caption{SDSS $gri$ images of the TDE host galaxies. Images are 30\arcsec$\times$30\arcsec. If these galaxies have experienced a recent merger, they are already in the coalescence phase.}
\label{fig:ps}
\end{figure}

If the starburst was triggered by a galaxy-galaxy merger, one possible mechanism that could boost the TDE rate after the starburst is a supermassive black hole (SMBH) binary. The age since the starburst began constrains the galaxy merger mass ratio, assuming that the starburst coincided with the coalescence of the two galaxies. The SDSS images of the TDE host galaxies do not show a an obvious separation into two merging components (see Figure \ref{fig:ps}), so if the galaxies have recently interacted, they have already begun to coalesce. In an unequal mass merger, the dynamical friction timescale for the smaller galaxy to fall into the larger one depends on the mass ratio \citep{Taffoni2003}. In Figure \ref{fig:age}b, we plot the age since the starburst began, with the top x-axis showing the corresponding merger mass ratio derived from the dynamical timescale \citep{Taffoni2003}. These merger ratios are the most unequal possible to have coalesced by the current starburst age. We assume the most conservative case considered by \citet{Taffoni2003}, so many galaxy mergers with these mass ratios would have longer dynamical fraction timescales. All of the hosts but that of PTF09axc are consistent with a merger mass ratio more equal than $12:1$.  

The galaxy mass ratios should be comparable to the SMBH mass ratio formed in the merger, as SMBH mass correlates with galaxy bulge mass for a variety of galaxy types \citep[e.g.,][]{McConnell2012}.  \citet{McConnell2012} find the scatter in the SMBH mass - bulge mass relation to be 0.3-0.4 dex. We estimate the scatter in the bulge mass - stellar mass relation (0.56 dex) using the catalog by \citet{Mendel2013}, and only considering galaxies with log stellar mass $<$10.5 to be consistent with the possible merger progenitors of the TDE host galaxies. Adding these uncertainties in quadrature produces an uncertainty in the total stellar mass to SMBH mass relation of 0.69 dex. 

\citet{Chen2011} predict that the TDE rate scales with the SMBH mass ratio $q = M_{secondary}/M_{primary}$ as $N_* \propto q^\alpha$, where $N_*$ is the number of TDEs during the SMBH binary coalescence, $\alpha = (2-\gamma)/(6-2\gamma)$, and $\gamma$ is the slope of the inner stellar density profile $\rho_* \propto r^{-\gamma}$. If $\gamma$ ranges from 1.5-2, as in \citet{Chen2011}, the dependence of the number of TDEs on $q$ is weak, with $\alpha = 0-0.16$. In this case, we would expect to see many more TDEs in minor mergers (lower $q$, or more unequal), because minor mergers are more common than major mergers.  For example, simulations predict that the distribution of merger mass ratios is $\propto q^{-0.3} (1-q)^{1.1}$ \citep{Stewart2008}. If the TDE rate were driven by SMBH binaries, TDEs would be primarily seen after minor mergers, unless the inner density profile is exceptionally flat ($\gamma < 0.5$). However, the inner density profile in post-starburst galaxies can be quite steep and may itself be driving the enhanced TDE rate \citep{Stone2016}. The preference of TDEs for post-merger galaxies with mass ratios more equal than $12:1$ is thus inconsistent with arising from SMBH binary interactions, if the TDE rate does scale weakly with the SMBH mass ratio \citep{Chen2011}.

\subsection{Stellar Masses of TDE Host Galaxies}
\label{sec:stmass}

We determine stellar masses of the TDE host galaxies as part of our stellar population fitting methodology. These stellar masses are more accurate than those derived assuming a single stellar population, as a recent starburst will impact the mass-to-light ratio. The derived stellar masses range from $10^{9.4} - 10^{10.3}$ M$_\sun$ (Table \ref{table:sfh2}). The range in stellar masses is connected that of the supermassive black hole masses, through the black hole -- bulge relation. To estimate bulge masses (also shown in Table \ref{table:sfh2}), we use the bulge to total mass ratios from \citet{Mendel2013}, for the four galaxies bright enough to have bulge:disk decompositions. We caution that bulge:disk decompositions can be unreliable for post-starburst galaxies, due to the bright, centrally concentrated A star population. We use the relation from \citet{McConnell2012} to estimate black hole masses from the bulge masses. The black hole masses range from $10^{5.5} - 10^{7.5}$M$_\sun$. These are consistent with the black hole masses expected of UV/optical TDEs \citep{Stone2015, Kochanek2016} and with an upper limit of $10^8$M$_\sun$, above which the tidal radius is inside the event horizon and the TDE undetectable. Given these black hole masses, we also note the stellar progenitor masses predicted by \citet{Kochanek2016} are consistent with our constraints in \S\ref{sec:startiming}.

The stellar masses of the TDE hosts are low compared to the quiescent Balmer-strong galaxies from the SDSS, where 68\% have stellar masses $10^{10.1} - 10^{11.2}$M$_\sun$. However, lower mass post-starbursts are too faint to be included in the SDSS spectroscopic survey at the redshifts of many of the TDE hosts. Therefore, we must compare the absolute $r$-band magnitudes of the TDE hosts to the volume corrected $r$-band galaxy luminosity function of the SDSS quiescent Balmer-strong galaxies. We calculate the volume corrections as in \citet{Simard2011}. The volume corrections appropriate for the TDE host galaxies is unknown, so we crudely compare two bins in $M_r$: [-23, -21] mag and [-21, -18] mag. All eight TDE hosts are in the fainter bin. The ratio of the volume-corrected number of galaxies in the first bin to the second bin is 0.13, well within the binomial confidence error for the TDE host sample. Thus, if the TDE hosts have a luminosity function similar to the quiescent Balmer-strong sample, the lack of TDEs observed in intrinsically bright hosts is due to the low numbers of these bright galaxies. 

With a theoretical upper limit of $10^8$M$_\sun$ on the black hole mass of the TDE host galaxies, we expect to see host galaxy stellar masses of up to $\sim 10^{10.8}$M$_\sun$. With the current low number of TDEs, it is unclear whether the observed upper bound on TDE host stellar mass of $10^{10.3}$ M$_\sun$ arises primarily from the falling galaxy stellar mass function, a preference of TDEs for lower stellar mass galaxies, or in fact from the limit on the black hole mass for which TDEs would be observable. With more events, a galaxy stellar mass function can be derived for the TDE hosts to search for any deviations from that expected for the quiescent Balmer-strong galaxies.

\subsection{Hidden Emission Line Ratios}

\begin{figure}
\includegraphics[width=0.5\textwidth]{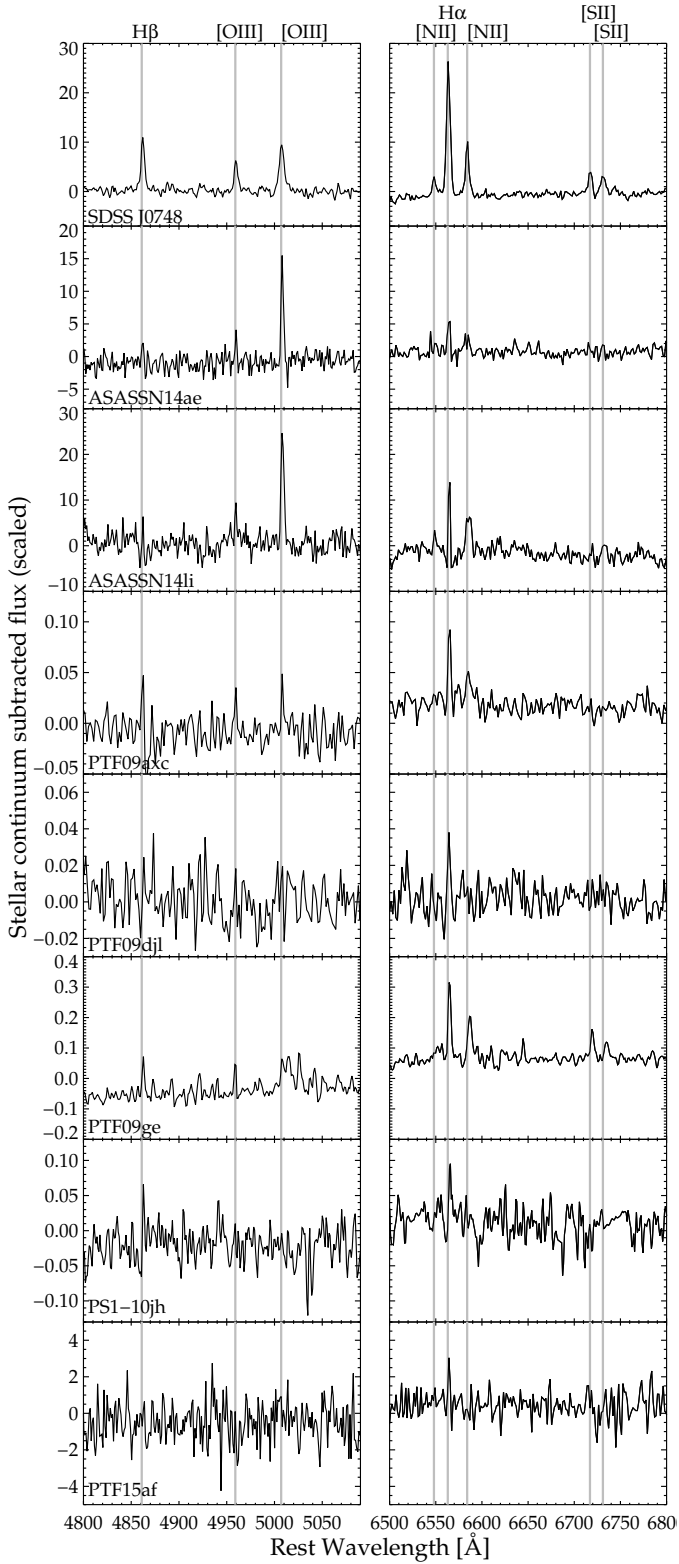}
\caption{Residual TDE host galaxy spectra after subtraction of the best fit stellar population model. This corrects for the strong Balmer absorption, and uncovers hidden line emission (see Table \ref{table:sfh2}).}
\label{fig:spec}
\end{figure}

For TDE host galaxies, weak emission lines may be hidden in the strong absorption. Using the stellar population models, we can model the absorption masking the Balmer emission lines to disentangle the emission and absorption features. The emission line ratios [NII]$\lambda$6583/H$\alpha$ and [OIII]$\lambda$5007/H$\beta$ are shown in Table \ref{table:sfh2}. To calculate the flux from each emission line, we subtract the stellar population best fit model from the data spectrum. The residual spectra are plotted in Figure \ref{fig:spec}. We additionally correct the zero point of the continuum by subtracting off the median level of the surrounding spectrum ($\lambda$ 6400 - 6700 \AA, $\lambda$  4800 - 5100 \AA\ for each group of lines), excluding the line windows. For each line, we use the line windows defined by the MPA-JHU group \citep{Brinchmann2004, Tremonti2004}. The width of each line window is 20\AA. For three of the galaxies (hosts of ASASSN-14ae, ASASSN-14li, and PTF15af), we also have the line fluxes fit by the MPA-JHU group. Our results and theirs are consistent within the measurement errors in these cases, and we choose the fit results with smaller errors (MPA-JHU line ratios for ASASSN-14ae and ASASSN-14li, our line ratios for PTF15af). For the host galaxy of PTF09ge, the [OIII]$\lambda$5007 line is contaminated, and we estimate its true flux by measuring the [OIII]$\lambda$4959 line and assuming a ratio of  [OIII]$\lambda$4959/[OIII]$\lambda$5007$ = 1/3$.

\begin{figure*}
\includegraphics[width=0.5\textwidth]{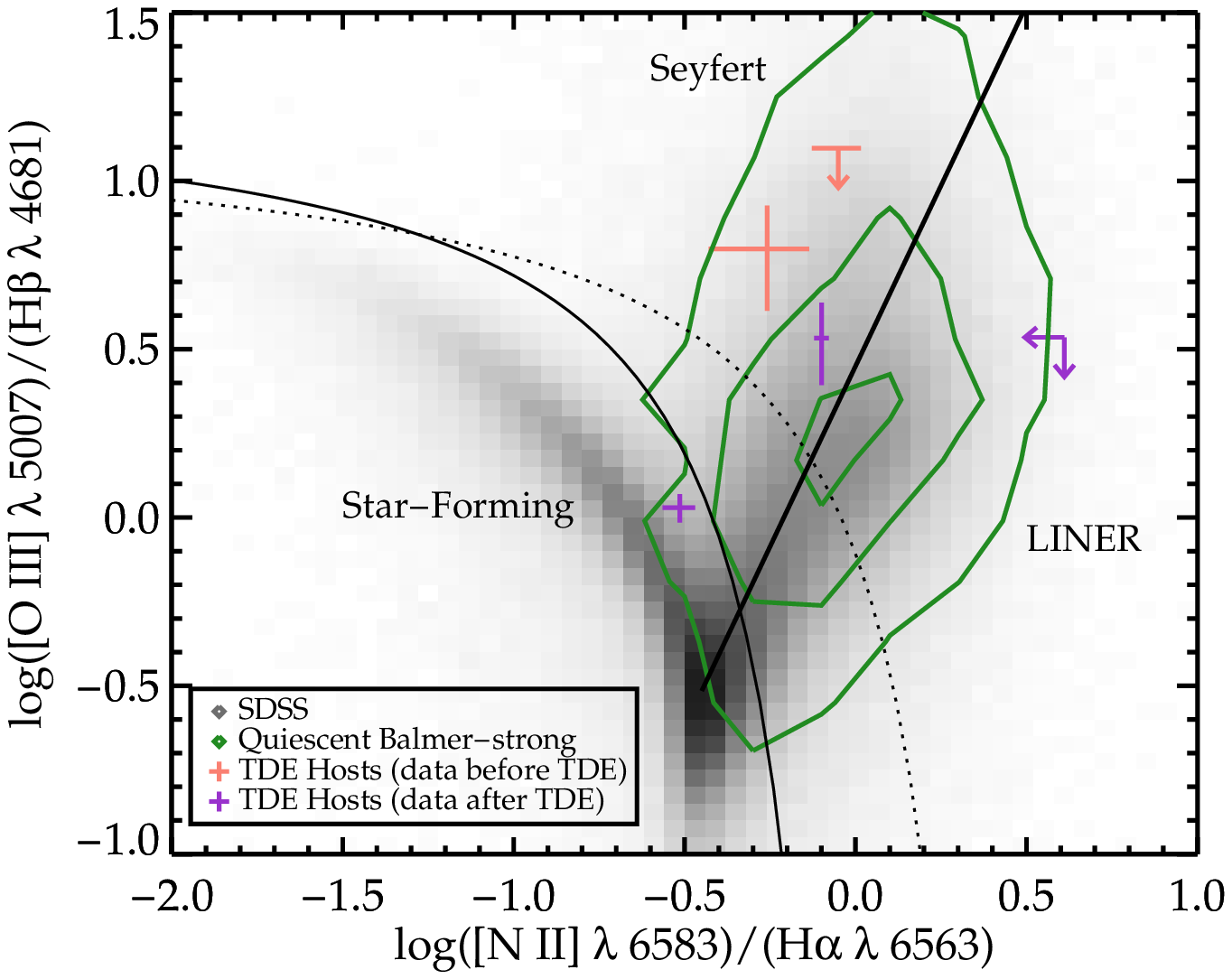}
\includegraphics[width=0.5\textwidth]{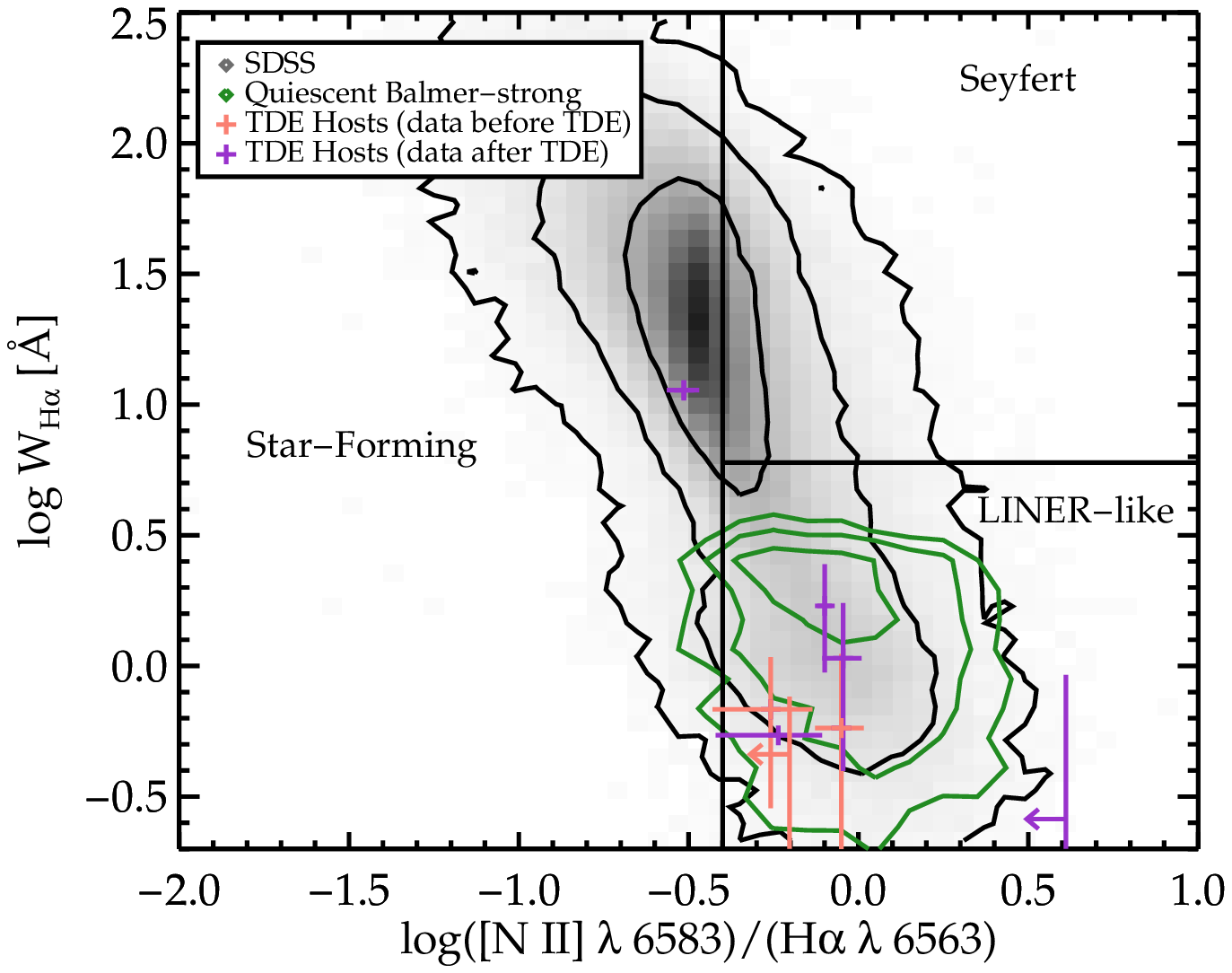}
\caption{{\bf Left:} BPT diagram for SDSS parent sample (shaded black), quiescent Balmer-strong galaxies with 9.5$<log[$M$_\star/$M$_\odot]<10.5$ (dark green contours: 20, 68, 95\%), and TDE host galaxies (orange and purple crosses). The TDE host galaxies are consistent with the quiescent Balmer-strong comparison sample, with most having emission line ratios inconsistent with star formation. We overplot the \citet{Kewley2001} and \citet{Kauffmann2003b} separation lines as dotted and solid lines respectively. {\bf Right:} WHAN diagram for SDSS parent sample, quiescent Balmer-strong galaxies with 9.5$<log[$M$_\star/$M$_\odot]<10.5$ (dark green contours: 30, 68, 85\%), and TDE host galaxies. All but the SDSS J0748 host are in the LINER-like region of this diagram. The TDE hosts lie offset from the centroid of the quiescent Balmer-strong comparison sample, especially when considering only the three TDE hosts with spectra from before the TDE (shown in orange), which have no possible contamination from the TDE.}
\label{fig:bpt}
\end{figure*}

We plot the emission line ratios for the TDE host galaxies on a BPT \citep{Baldwin1981} diagram in Figure \ref{fig:bpt}. The host of SDSS J0748 is in the star-forming region, the hosts of ASASSN-14ae and PTF09ge are in the Seyfert region, the host of ASASSN-14li could be in the Seyfert or LINER regions, and the host of PTF09djl is largely unconstrained. We compare the BPT location of the TDE host galaxies to the quiescent Balmer-strong galaxies from SDSS with similar stellar masses, 9.5$<log[$M$_\star/$M$_\sun]<10.5$, and find them consistent. Like the SDSS quiescent Balmer-strong galaxies, the TDE hosts often have emission line ratios inconsistent with ionization from star formation. We note that the high energy TDE Swift J1644 has a host galaxy with emission line ratios found to be consistent with star formation \citep{Levan2011}.

There are several possible ionization sources in these galaxies, which might be related to the mechanism boosting the TDE rate. One possibility is a low luminosity AGN fueled by a circumnuclear gas reservoir. Interactions with the gas disk can increase the TDE rate by a factor of 10$\times$ \citep{Kennedy2016}. However, in a recent merger or starburst, merger shocks \citep{Rich2015, Alatalo2016}, which may persist through the post-starburst phase, and post-AGB stars \citep{Yan2012} also can produce emission line ratios in the LINER/Seyfert portions of the BPT diagram. Spatially resolved spectroscopy, such as that in \citet{Prieto2016}, combined with stellar population modeling to account for the strong Balmer absorption, is needed to differentiate among these possibilities. 

We plot the TDE host galaxies on a WHAN \citep{CidFernandes2010} diagram in Figure \ref{fig:bpt}. In comparison to the BPT diagram, the lack of a required H$\beta$ detection allows for more host galaxies to be considered. With the exception of the host of SDSS J0748, all galaxies lie in the LINER-like portion of this diagram. It is not surprising that some of the SDSS quiescent Balmer-strong and TDE host galaxies lie in the ``Seyfert" portion of the BPT diagram, yet are in the ``LINER-like" portion of the WHAN diagram, as this is common for galaxies with weak emission lines \citep{CidFernandes2010}. In the WHAN diagram, unlike in the BPT diagram, the TDE hosts lie offset from the centroid of the comparison sample of SDSS quiescent Balmer-strong galaxies. We note that the three TDE hosts with spectra from before the TDE, and thus not contaminated by the recent TDE, are more offset from the comparison galaxies. Their offset in H$\alpha$ EW is significant at 3$\sigma$, and that in [NII]/H$\alpha$ at 2$\sigma$, suggesting a lower electron density, a softer radiation field \citep{Kewley2013}, or a lower residual star formation rate. However, we caution that these conclusions rely on only three host galaxies.

The host galaxy of SDSS J0748 is an outlier in many of these comparisons. It is the only host galaxy with significant current star formation. This may not be surprising, as its TDE was detected in a different manner than the rest, serendipitously from the SDSS \citep{Wang2011} during a search for narrow high ionization coronal emission lines. 

\section{Conclusions}

We fit stellar population models to UV and optical photometry and optical spectroscopy of eight host galaxies of optical/UV-detected tidal disruption events (TDEs). We determine the duration of the recent star formation episode, the time elapsed since it ended, and the fraction of stellar mass produced, breaking the degeneracy in these quantities. We also determine the stellar mass of the galaxies and measure residual emission line ratios in their model-subtracted spectra. We compare the TDE host galaxy properties to other quiescent galaxies with strong Balmer absorption and with the general galaxy population. We conclude the following:

\begin{enumerate}
\item Most (6/8) of the TDE hosts have short (25$-$200 Myr) periods of star formation, consistent with a recent starburst rather than a long-term decline in star formation. The eight TDE host galaxies thus consist of six post-starburst galaxies, one star-forming galaxy, and one quiescent galaxy which experienced a long period of recent star formation.

\item Most (6/8) TDE host galaxies have post-burst ages of $60 - 600$ Myr. This range is physical, exceeding our measurement errors, and indicates that TDEs do not occur at a specific time after the starburst ends. 

\item With the post-burst ages, we can constrain the mass of the disrupted star, assuming it formed in the burst or before. The range in post-burst ages is much larger than the expected post-main sequence evolution timescales, implying that a specific phase of post-main sequence evolution is unlikely to be the cause of the enhanced TDE rate after the starburst. If the disrupted star was a main sequence star, the post-burst ages constrain the upper limits on its mass to be $\sim$2.5, $\sim$4, $\sim$6, and $\sim$9 M$_\sun$ for the seven non-starforming hosts. In other words, O stars, as well as most B and massive A stars, are excluded as TDE progenitors.

\item If the starburst arose from a galaxy-galaxy merger, the time elapsed since the starburst began constrains the coalescence timescale and thus limits the merger mass ratio to more equal than 12:1 in most (7/8) TDE hosts. This ratio is unusual, as more unequal galaxy mergers are more common. If this ratio also reflects that of the central supermassive black hole binary, it disfavors the scenario in which the TDE rate is boosted by the binary in a way that is insensitive to its mass ratio \citep[e.g.,][]{Chen2011}.

\item The fraction of stellar mass created in the burst is $0.5 - 10$\% for most (7/8) of the TDE hosts. These burst mass fractions do not generate enough stars compared to those formed by a typical star-forming galaxy over the same time to account for the  $30-200\times$ boost in TDE ranges. If simply adding more stars does not explain the rate enhancement, their concentration in the core must be more important. Future work is needed to assess the spatial distribution of these newly formed stars in the core.

\item The TDE host galaxies have stellar masses $10^{9.4} - 10^{10.3}$, consistent with the SDSS volume-corrected distribution of post-starbursts. Using bulge:disk decompositions from \citet{Mendel2013} and the black hole -- bulge relation from \citet{McConnell2012}, we infer black hole masses of $10^{5.5} - 10^{7.5}$M$_\sun$. These are consistent with the black hole masses expected for UV/optical TDEs \citep{Stone2015, Kochanek2016} and with an upper limit of $10^8$M$_\sun$, above which the TDE would be hidden within the event horizon. With the current low number of observed TDEs, it is unclear whether the upper bound on TDE host stellar mass is primarily driven by the falloff in the galaxy stellar mass function at high mass, by a preference of TDEs for lower stellar mass galaxies, or in fact by the upper limit on the black hole mass for which TDEs would be observable.

\item The TDE host galaxies that can be placed on a BPT \citep{Baldwin1981} diagram have ionization sources inconsistent with star formation, except for the star-forming host of SDSS J0748. Their distribution is consistent with quiescent Balmer-strong galaxies from the SDSS at comparable stellar masses. The residual emission line ratios uncovered here point to shocks, post-AGB stars, or a low-luminosity AGN as possible ionization sources. In the case of an AGN, circumnuclear gas accreting onto the SMBH could boost the TDE rate \citep{Kennedy2016}.

\item On the WHAN \citep{CidFernandes2010} diagram, at least 5/8 TDE host galaxies lie in the LINER-like region. The three TDE host galaxies with data from before the TDE (and thus uncontaminated by it) are offset to lower H$\alpha$ EW and [NII]/H$\alpha$ than the quiescent Balmer-strong galaxies from the SDSS at comparable stellar masses. This may indicate a lower electron density, a softer radiation field, or decreased levels of residual star formation in these TDE host galaxies.

\end{enumerate}

\acknowledgements
We thank Nick Stone and Sjoert van Velzen for useful discussions. KDF acknowledges support from NSF grant DGE-1143953, P.E.O., and the ARCS Phoenix Chapter and Burton Family.  Support for IA was provided by NASA through the Einstein Fellowship Program, grant PF6-170148. AIZ acknowledges funding from NSF grant AST-0908280 and NASA grant ADP-NNX10AE88G. 

Funding for SDSS-III has been provided by the Alfred P. Sloan Foundation, the Participating Institutions, the National Science Foundation, and the U.S. Department of Energy Office of Science. The SDSS-III web site is http://www.sdss3.org/. SDSS-III is managed by the Astrophysical Research Consortium for the Participating Institutions of the SDSS-III Collaboration including the University of Arizona, the Brazilian Participation Group, Brookhaven National Laboratory, Carnegie Mellon University, University of Florida, the French Participation Group, the German Participation Group, Harvard University, the Instituto de Astrofisica de Canarias, the Michigan State/Notre Dame/JINA Participation Group, Johns Hopkins University, Lawrence Berkeley National Laboratory, Max Planck Institute for Astrophysics, Max Planck Institute for Extraterrestrial Physics, New Mexico State University, New York University, Ohio State University, Pennsylvania State University, University of Portsmouth, Princeton University, the Spanish Participation Group, University of Tokyo, University of Utah, Vanderbilt University, University of Virginia, University of Washington, and Yale University.

\bibliographystyle{apj}
\bibliography{earefs.bib}

\end{document}